\documentclass[twocolumn,english,aps,prb,floatfix]{revtex4}
\usepackage{amssymb}
\usepackage[dvips]{graphicx}

\begin{document}

\title{Ultrafast Conductivity Dynamics in Pentacene Probed using Terahertz Spectroscopy}
\author{V.~K.~Thorsm{\o}lle}
\email{vthorsmolle@lanl.gov}
\author{R.~D.~Averitt}
\author{X.~Chi}
\author{D.~J.~Hilton}
\author{D.~L.~Smith}
\author{A.~P.~Ramirez}
\author{A.~J.~Taylor}
\address{Los Alamos National Laboratory, Los Alamos, NM 87545.}

\begin{abstract}
We present measurements of the transient photoconductivity in pentacene single crystals using optical-pump THz-probe spectroscopy. We have measured the temperature and fluence dependence of the mobility of the photoexcited charge carriers with picosecond resolution. The pentacene crystals were excited at 3.0 eV which is above the bandgap of $\sim$2.2 eV and the induced change in the far-infrared transmission was measured. At 30 K, the carrier mobility is $\mu$ $\approx$ 0.4 cm$^2$/Vs and decreases to $\mu$ $\approx$ 0.2 cm$^2$/Vs at room temperature. The transient terahertz signal reveals the presence of free carriers that are trapped on the timescale of a few ps or less, possibly through the formation of excitons, small polarons, or trapping by impurities.
\end{abstract}
\pacs{78.47.+p,72.20.-i,72.80.Le}

\maketitle

The promise of organic semiconductors for technological
applications in electronics and photonics has spurred an immense
interest in these materials \cite{BaldoNature}. However, despite
intense research efforts in recent years, the nature of charge
transport and photoexcitations in conjugated polymers and organic
molecular crystals is still not well understood and remains
controversial \cite{HegmanCanada}.

Conductivities in undoped organic semiconducters are orders of
magnitude smaller than in inorganic semiconducters such as Si and
GaAs. Polyacenes such as naphthalene (Nph) and pentacene (Pc) are
conjugated molecules with mobilities, $\mu$, on the order of 1
cm$^2$/Vs. The main contributing factors to such small mobilities
are thought to be related to weak intermolecular interactions.
These factors include strong localization, polaron formation and
trapping by impurities. However, there is little firm experimental
evidence for the importance of any single one of these factors in
single crystals. The temperature $(T)$ dependence of the mobility
in these crystals is observed to increase as the temperature is
lowered \cite{WartaStehleKarl}. This has been attributed to a
transition from a polaron hopping transport to a band-like
transport of carriers at lower temperatures described by some
theoretical models \cite{Silinsh,Silinsh2} and seen by Schlein
{\sl et al.} in Nph \cite{Schlein}. In some materials the
temperature dependence follows $\mu\propto T^{-n}$, $n>0$. Warta
{\sl et al.} observed this behavior in Nph until low temperature,
$\sim$30 K, where it then levels off at values greater than 100
cm$^2$/Vs \cite{WartaStehleKarl,WartaKarl}. Clearly, the nature of
charge carrier transport is still not completely understood.

It is not clear either whether electronic excitations are best
described by the molecular exciton model, or by the semiconductor
band model \cite{Sariciftci}. In the former case, the excited
states are localized and the primary photoexcitations are
excitons, which can dissociate into free polarons. In the latter
case, they are delocalized and mobile polarons are created
directly from free electron-hole pairs by the absorption of light.
Ultrafast optical measurements are important in this regard in
that the dynamics of the photogenerated carriers can be temporally
resolved. While the optical-pump probe technique is a noncontact
technique and yields important information of the relaxation
dynamics, optical-pump terahertz-probe is a direct probe of the
nonequilibrium carrier conductivity dynamics. Terahertz
time-domain spectroscopy (THz-TDS) without an optical pump is an
ultrafast optical technique in which near single-cycle freespace
electric field transients of a duration of about 1 ps and spectral
width on the order of 1 THz are used to measure the complex
conductivity of the sample. This is a coherent technique, and by
varying the probe delay time with respect to an optical excitation
pulse the induced conductivity changes of the sample can be
measured with sub-picosecond (sub-ps) resolution. Thus, THz-TDS
has become an important technique in the study of the far-infrared
conductivity in condensed matter systems ranging from carrier
transport in semiconductors to the dynamics of fluids
\cite{Thorsmolle,Rich,Heinz,Schmuttenmaer,Lui,Averitt1,Grischkowsky,Hegmann,Keiding}.

\begin{figure}[h]
\includegraphics{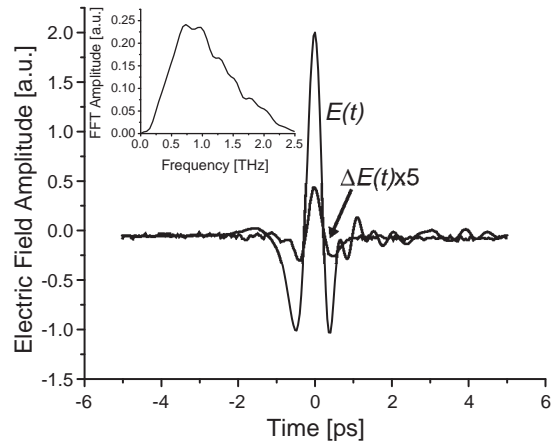}
\caption{Electric field $E(t)$ of THz pulse transmitted through Pc
crystal, and induced change in electric field $\Delta E(t)$ at $T
= 4$ K. The inset shows the fast Fourier transformed amplitude
spectrum of $E(t)$.\label{fig: THz pulses}}
\end{figure}

Here we present optical-pump THz-probe conductivity measurements
on Pc single crystals. High quality Pc single crystals were grown
in a flow of inert gas. The crystals used were typically 3 mm x 3
mm and approximately 50 $\mu$m thick. The experiments utilized a
commercial-based regeneratively amplified Ti:Al$_2$O$_3$ system
operating at 1 KHz producing nominally 2.0 mJ, sub-50 fs pulses at
1.5 eV. The THz pulses were generated and detected using
electrooptic techniques. The schematic of the THz-setup is shown
in Ref.~\cite{Averitt2}. The Pc crystals were here excited at 3.0
eV which is above the bandgap of ~2.2 eV. The experiments were
performed in transmission with the electric field of the THz
pulses in the $ab$-plane. The diameter of the THz beam is
apertured to 2 mm at the sample position, and the diameter of the
pump beam is $\sim$3 mm. The samples were mounted inside an
optical He cryostat. The results presented are obtained from
several samples.

Figure \ref{fig: THz pulses} shows the electric field of the THz
pulse transmitted through a Pc crystal without optical excitation,
compared to the induced change in the electric field with optical
excitation. There is a decrease in the transmitted electric field
associated with the conductivity of the mobile carriers (i.e.
Drude-like response). No shift in the phase of the induced change
of the THz pulse is observed. Thus it is valid to measure changes
in the peak amplitude of the transmitted THz pulses to determine
the photoinduced conductivity (i.e. $\Delta\sigma\propto \Delta
E/E)$.

\begin{figure}[h]
{\includegraphics{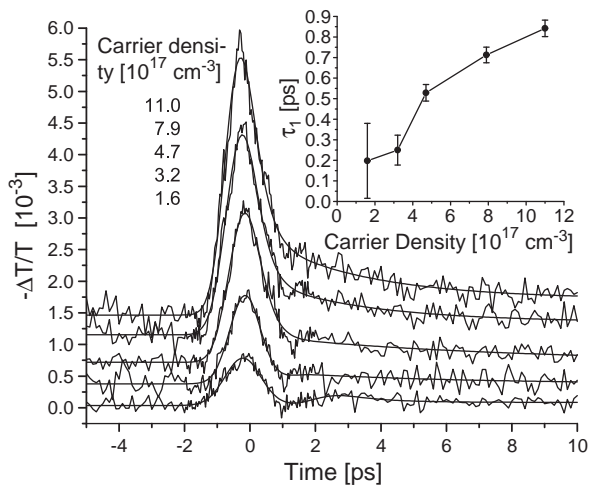}} \caption{Differential transmission of
the peak of the THz probe pulse versus pump-probe delay time at
various fluences at $T = 20$ K. The plots are displaced vertically
for clarity. The inset shows the fast exponential decay time
versus pump fluence.\label{fig: Fluence dependence}}
\end{figure}

Figure \ref{fig: Fluence dependence} shows the induced change in
transmission of the THz electric field $-\Delta$T/T at incident
pump fluences ranging from 0.12 to 0.83 mJ/cm$^2$ producing
carrier densities from $1.6\times10^{17}$ to 1.1$\times$10$^{18}$
cm$^{-3}$ at $T = 20$ K. The photoconductivity transient increases
with increasing fluence indicating an increasing conductivity with
increasing carrier density. The rise time is resolution limited by
the THz setup. The initial exponential decay increases with
increasing fluence (see inset to Figure \ref{fig: Fluence
dependence}), indicating some interaction between the carriers
creating a bottleneck effect. This behavior could also simply be
due to a saturation of the trap density where carriers are trapped
by defects. At higher fluences, an additional exponential
relaxation component develops with a longer lifetime leveling off
at $\sim$4 ps. $-\Delta$T/T is close to linear with pump fluence.

\begin{figure}[h]
{\includegraphics{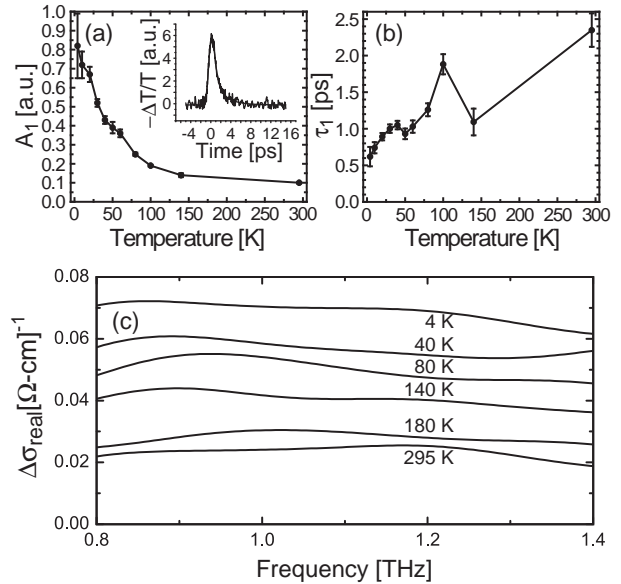}} \caption{Temperature dependence of
various parameters at a fluence of $I=0.21$ mJ/cm$^2$. (a) Fast
decay amplitude versus temperature. The inset shows $-\Delta$T/T
at $T = 4$ K. (b) Fast decay time versus temperature. (c) Induced
change in real conductivity versus frequency at various
temperatures. \label{fig: Temperature Dependence}}
\end{figure}

\begin{figure}[h]
\includegraphics{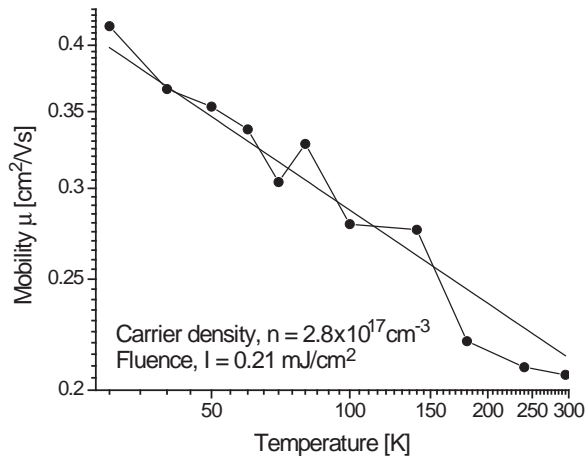}
\caption{Photoinduced carrier mobility versus temperature. The
straight line is a fit to $T^{-n}$, where $n=0.27$.
\label{fig:mobility}}
\end{figure}

The differential transmission versus temperature displays a
similar behavior where the peak of the transient signal, and hence
the transient conductivity, increases as the temperature is
lowered. We attribute this to an increase in carrier mobility.
Figure \ref{fig: Temperature Dependence}(a) shows the fast decay
amplitude to be increasing with decreasing temperatures, following
the mobility (see Figure \ref{fig:mobility}). The fast decay time
shown in Figure \ref{fig: Temperature Dependence}(b) is seen to
increase with increasing temperature which is not immediately
obvious, but might be due to thermal population of exciton levels
or trapped states with higher temperatures. Figure \ref{fig:
Temperature Dependence}(c) shows the induced change in the real
conductivity, and it is seen to be increasing with decreasing
temperatures. The induced change in the imaginary conductivity is
close to zero. Thus, it was not possible to fit the data to a
Drude model with such small induced conductivities.

Figure \ref{fig:mobility} shows the carrier mobility versus
temperature obtained directly from the time domain data,
$-\Delta$T/T, and the induced increase in conductivity,
$\Delta\sigma=ne\mu$ \cite{Lui}. Here, $n$ is the carrier density,
$e$ is the electronic charge, and $\mu$ is the effective carrier
mobility. $\Delta\sigma$ is obtained from the approximation for a
thin film on a semi-infinite insulating media given by

\begin{eqnarray}
\Delta\sigma & \cong & -\frac{\Delta\mathrm{T}}{\mathrm{T}}\frac{(1+N)}{Z_o d},
\end{eqnarray}

\noindent

where $N=1.6$ is the refractive index of the media (unpumped
crystal), $d=14$ $\mu$m is the optical penetration depth at the
pump wavelength, and $Z_o=377$ $\Omega$ is the impedance of free
space. The carrier density decays exponentially by $d$, but is
assumed constant over the excitation thickness.

At low temperatures, $T = 30$ K, the photoinduced carrier mobility
is $\mu\approx$ 0.4 cm$^2$/Vs and $\mu$ decreases to $\mu\approx$
0.2 cm$^2$/Vs at room temperature. These values for the mobility
assumes a 100$\%$ internal quantum efficiency conversion to free
carriers, which may be an overestimate of the carrier density.
These results provide a lower limit on the mobility. These values
are comparable to values found in Pc crystals by Bukto {\sl et
al.} \cite{Butko} in a field effect geometry, as well as by
Hegmann {\sl et al.} also using the same technique of optical-pump
THz-probe spectroscopy on functionalized Pc crystals
\cite{Hegmann}. The temperature dependence of the mobility shown
in Figure \ref{fig:mobility} follows the $T^{-n}$ power-law in the
temperature interval 30$-$300 K, where $n=0.27\pm 0.05$. The low
value of $n$ does not give a clear indication of a pure band-like
transport at low tempertures, and the continuation of the typical
power-law dependence into the high-temperature regime is not
completely understood \cite{Silinsh}. The continuation of the
power-law dependence into the high-temperature regime was also
observed in Nph by Warta {\sl et al.} \cite{Schlein,WartaKarl}.
The fact that the conductivity shown in Figure \ref{fig:
Temperature Dependence}(c) is seen to be increasing with
decreasing temperature does however indicate a band-like
transport.

The fast rise time for $-\Delta$T/T of $\sim$0.5 ps suggests free
carriers or polarons to be photoexcited directly, because THz-TDS
is sensitive to the presence of free charge carriers. This rise
time is comparable to the response time of the THz setup, and it
is therefore likely that the formation of mobile charge carriers
occur on a shorter timescale. This same behavior was observed by
Hegmann {\sl et al.} in functionalized Pc \cite{Hegmann}. Another
group have reported the formation of polarons within $\sim$100 fs
in a conjugated polymer (PPV thin films) \cite{Moses,Miranda},
thus supporting the semiconductor band model. Yet another group
finds that polarons are first generated by electric-field-assisted
dissociation of primary excitons on a timescale of 10 ps in light
emitting diodes ($m$-LPPP thin films) \cite{Grauper}, thus
supporting the molecular exciton model. Clearly, further studies
are needed to address this controversy.

In conclusion, we have measured the transient photoconductivity in
Pc single crystals using THz-TDS. The rapid onset of
photoconductivity suggests the primary photoexcitations to be
mobile charge carriers. The photoinduced response consists of a
two-component exponential relaxation. The initial fast component
relaxes within a few ps or less, possibly through the formation of
excitons, small polarons, or trapping by impurities. Photoinduced
mobilities of at least $\mu$ = 0.4 cm$^2$/Vs at 30 K was obtained.
The temperature dependence of the photoinduced mobilities is found
to follow a power-law, $T^{-n}$, where $n=0.27$.

This research was supported by the Los Alamos Directed Research and Development Program of the U.S. Department of Energy.

\end{document}